\documentclass[a4paper,11pt]{article}

\usepackage{contribution}

\newcommand{\weblink}[2][]{%
    \ifthenelse{\equal{#1}{}}%
    {\textnormal{\url{#2}}}%
    {\textnormal{\href{#2}{#1}}}%
}

\newcommand{\acknowledgements}[1]{%
  \bigskip\bigskip
  \textsf{\textbf{\Large Acknowledgements}} \\[2ex]
  {#1}
  \bigskip
}

\def\beq{\begin{equation}}
\def\eeq#1{\label{#1}\end{equation}}
\def\eeqn{\end{equation}}

\def\beqa{\begin{eqnarray}}
\def\eeqa#1{\label{#1}\end{eqnarray}}
\def\eeqan{\end{eqnarray}}

\let\bar=\overbar

\def\Dslash{\not{\hbox{\kern-4pt $D$}}}
\def\dslash{\not{\hbox{\kern-2pt $\del$}}}

\def\msb{{\bar{\ssstyle M \kern -1pt S}}}

\newcommand{\contribution}[7][]{%
  \clearpage
  \thispagestyle{plain}
  \ifthenelse{\equal{#1}{}}
  {\hypersetup{pdftitle={#2}}}
  {\hypersetup{pdftitle={#1}}}
  \hypersetup{pdfauthor={{#3} {#4}}}
  {\centering\normalfont\LARGE\bfseries\sffamily #2 \par\nobreak}
  \lhead{}
  \chead{%
    \textit{\footnotesize XIV International Conference on Hadron Spectroscopy
      (\weblink[\textit{hadron2011}]{http://www.hadron2011.de}), 13-17 June 2011, Munich, Germany}%
  }
  \rhead{}
  \bigskip
  \begin{center}
    {#3} {#4}\ifthenelse{\equal{#6}{}}{}{\footnote{\weblink[#6]{mailto:#6}}}
    \ifthenelse{\equal{#7}{}}{}{#7} \\
    \textit{#5}
  \end{center}
  \bigskip
}

\renewcommand{\abstract}[1]{%
  \begin{center}
    \begin{minipage}{0.85\textwidth}
      \begin{footnotesize}
        #1
      \end{footnotesize}
    \end{minipage}
  \end{center}
  \bigskip
}

\begin{document}

%
%
%
%
%
{  

%

\contribution[]
{The $pp \to p \Lambda K^+ $ and $pp \to p \Sigma^0 K^+$ Reactions \\ in the Chiral Unitary Approach}
{Hua-Xing}{Chen}  
{$^a$Departamento de F\'{\i}sica Te\'orica and IFIC, Centro Mixto Universidad de Valencia-CSIC,
Institutos de Investigaci\'on de Paterna, Aptdo. 22085, 46071 Valencia, Spain \\
$^b$School of Physics and Nuclear Energy Engineering, Beihang University, Beijing 100191, China \\
$^c$Department of Physics, Zhengzhou University, Zhengzhou, Henan 450001, China }
{hxchen@ific.uv.es}
{\!\!$^{,a,b}$, Ju-Jun Xie$^{a,c}$, and E.Oset$^{a}$}
%

\abstract{%
We study the $pp \to p \Lambda K^+$ and $pp \to p \Sigma^0 K^+$ reactions near threshold by using a chiral unitary approach. We consider the single-pion and single-kaon exchange as well as the final state interactions of nucleon-hyperon, $K$-hyperon and $K$-nucleon systems. Our results on the total cross section of
the $pp \to p \Lambda K^+$ reaction is consistent with the experimental data, and the experimental observed strong suppression of $\Sigma^0$ production compared to $\Lambda$ production at the same excess energy can also be explained in our model.
}
%

\section{Introduction}

By using the chiral unitary approach, we study the $pp \to p \Lambda K^+$  and $pp \to p \Sigma^0 K^+$ reactions near
threshold considering pion and kaon exchanges~\cite{Xie:2011me}, where the $p \Lambda$ final state interaction (FSI) is very important~\cite{theory,Gasparian:1999jj}. The $\pi N \to K \Lambda$ amplitude also appears in this scheme, and the unitarization of this amplitude produces naturally the $N^*(1535)$ resonance~\cite{Kaiser}, such that we can make a quantitative statement on its relevance in the $pp \to p\Lambda K^+$ reaction. We find that the $p \Lambda$ interaction close to threshold is very strong~\cite{Dover}, and the FSI due to this source is unavoidable in an accurate calculation and we also take it into account.

We use a dynamical model similar to the one in Ref.~\cite{Gasparian:1999jj} but we allow all pairs in the final state to undergo FSI, as a consequence of which we obtain a contribution from the $N^*(1535)$ using chiral unitary amplitudes. Our approach also differs from the other approaches on how the FSI is implemented, and for this we follow the steps of Ref.~\cite{junkojuan}. Furthermore, the experimental total cross section for the $p p \to p \Sigma^0 K^+$ reaction is strongly suppressed compared to that of the $p p \to p \Lambda K^+$ reaction at the same excess energy. This was explained by a destructive interference between $\pi$ and $K$ exchange in the reaction $pp \to p \Sigma^0 K^+$~\cite{Gasparian:1999jj}.

\section{Formalism and ingredients}

\begin{figure}[htbp]
\begin{center}
\includegraphics[width=0.4\textwidth]{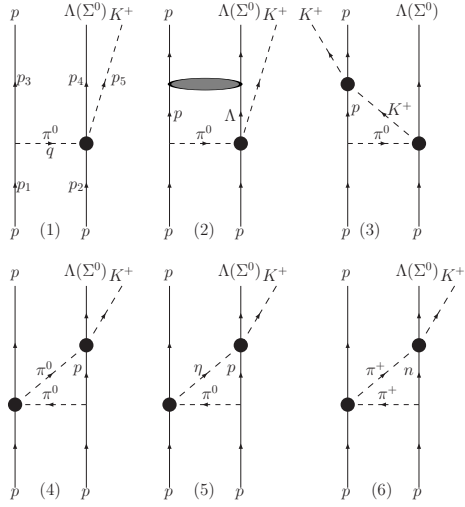}
\caption{The $\pi$ exchange mechanism of the $pp \to p \Lambda (\Sigma^0) K^+$ reactions including the final state interactions.}
\label{Fig:pidiagram}
\end{center}
\end{figure}

At the reaction threshold, the processes involving the exchange of $\pi$ and $K$ mesons are the dominant contributions, as in Ref.~\cite{Gasparian:1999jj} and other works of the Juelich group. Accordingly we show the dominant diagrams exchanging $\pi$ mesons in Fig.~\ref{Fig:pidiagram}, where the definitions of the kinematics ($p_1$, $p_2$, $p_3$, $p_4$, $p_5$, and $q$) are shown in the first diagram. Those exchanging $K$ mesons can be similarly obtained.
First we write out the amplitudes for elementary production processes. For the first diagram of Fig.~\ref{Fig:pidiagram}, we have
\begin{eqnarray}\label{eq:Api1}
{\cal A}^1_{\pi} &=& - F_{\pi NN}(q^2) f_{\pi^0 pp} \sigma_z(1) q_z
\frac{i}{q^2-m^2_{\pi}} T_{\pi^0 p \to K^+ \Lambda},
\end{eqnarray}
where $F_{\pi NN}(q^2)$ is the form factor containing a cutoff parameter $\Lambda_{\pi}$:
\begin{eqnarray}
F_{\pi NN}(q^2) &=& \frac{\Lambda^2_{\pi}-m_{\pi}^2}{\Lambda^2_{\pi}- q^2},
\end{eqnarray}

We can similarly obtain the ``elementary production amplitudes'' for the other diagrams, and the total production amplitude ${\cal M}$ can be written into two
parts:
\begin{eqnarray}
{\cal M} &=& {\cal M}_{\pi} + {\cal M}_{K}, \label{totalm}
\end{eqnarray}
where ${\cal M}_{\pi}$ is for those diagrams involving $\pi$ exchange ( ${\cal M}_{K}$ for $K$ exchange):
\begin{eqnarray}
{\cal M}_{\pi} &=& {\cal A}^1_{\pi} +\sum^6_{i=2} {\cal A}^i_{\pi}G_{\pi}^iT_{\pi}^i, \label{mpi}
\end{eqnarray}
where ${\cal A}^i_{\pi/K}$ are the elementary production processes which can be obtained similarly to Eq.~(\ref{eq:Api1}) and $G_{\pi}^i$ the loop functions of one meson and a baryon propagators, or two baryon propagators. Together with the final state interactions for meson-baryon cases (such as $T^3_{\pi} = T_{K^+ p \rightarrow K^+ p}$, etc.) and for baryon-baryon cases ($T^2_{\pi} = T_{\Lambda p \rightarrow \Lambda p}$, etc.), we can obtain the full total production amplitude ${\cal M}$.

The meson-baryon $G$-functions and $T$-matrices have been calculated in Refs.~\cite{osetKN}, and we only need to calculate the baryon-baryon ones which are done using the experimental data~\cite{ppdata}. We also consider the transition between $p p \to p \Lambda K^+$ and $p p \to p \Sigma^0 K^+$, which is discussed in Ref.~\cite{Xie:2011me} in detail.

\section{Numerical results and Discussion}

The total cross section versus the excess energy ($\varepsilon$) for the $pp \to p \Lambda K^+$ and $pp \to p \Sigma^0 K^+$ reactions are calculated by using a Monte Carlo multi-particle phase space integration program. The results for $\varepsilon$ from $0$ MeV to $14$ MeV is shown in Fig.~\ref{pltcs} for the $pp \to p \Lambda K^+$ reaction with the cutoff $\Lambda_{\pi}=1300$ MeV, together with the experimental data~\cite{ppdata} for comparison.
\begin{figure}[htbp]
\begin{center}
\includegraphics[width=0.4\textwidth]{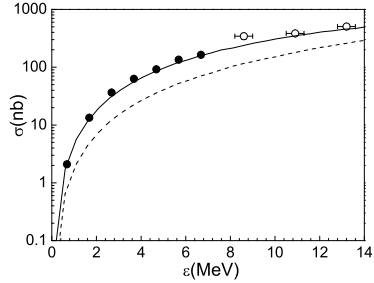}%
\caption{Total cross section vs excess energy $\varepsilon$ for the
$pp \to p \Lambda K^+$ reaction compared with experimental data from
Refs.~\cite{ppdata} (filled and open circles).} \label{pltcs}
\end{center}
\end{figure}
The solid and dashed lines show the results from our model with and without including the $p \Lambda$ FSI, respectively.

We can see that we can reproduce the experimental data quite well for the excess energy $\varepsilon$ lower than $14$ MeV. The dashed line is about two and a half times smaller than the experimental data at threshold but less than a factor of two smaller than experimental data at $\epsilon \sim 14$ MeV. This indicates that the $p\Lambda$ FSI is very important in the $pp \to p \Lambda K^+$ reaction close to threshold. This energy dependence of the FSI is what allows the determination of the $\Lambda N$ interaction in other approaches which do not try to get absolute cross sections~\cite{hinterask}.

\acknowledgements{%
This work is partly supported by DGICYT contracts No. FIS2006-03438,
FPA2007-62777, the Generalitat Valenciana in the program PROMETEO
and the EU Integrated Infrastructure Initiative Hadron Physics
Project under Grant Agreement No. 227431. Ju-Jun Xie acknowledges
Ministerio de Educaci\'{o}n Grant SAB2009-0116.
}


%

}  


\end{document}